\documentclass[a4paper,11pt]{article}
\usepackage{pos}

\usepackage{url}
\usepackage{nicefrac}
\usepackage{cuted}
\usepackage{slashed}
\usepackage{orcidlink}

\newcommand{\bp}{\bar{p}} % \bp = (p + p')/2
\newcommand{\thetaL}{\theta_\ell} % polar angle of muon's 3-momentum in leptons' CM frame
\newcommand{\phiL}{\phi_\ell} % azimuthal angle of muon's 3-momentum in leptons' CM frame
\newcommand{\M}{\mathcal{M}} % amplitude
 
\newcommand{\cffh}{\mathcal{H}} % CFF H
\newcommand{\cffe}{\mathcal{E}} % CFF E
\newcommand{\cffht}{\mathcal{\widetilde{H}}} % CFF Htilde
\newcommand{\cffet}{\mathcal{\widetilde{E}}} % CFF Etilde

\newcommand{\Jt}{\mathcal{J}^{(2)}}
\newcommand{\Jofplus}{\mathcal{J}^{(1, 5)+}}
\newcommand{\Jtfplus}{\mathcal{J}^{(2, 5)+}}

\newcommand{\thetaS}{\theta_S}
\newcommand{\phiS}{\phi_S}

\newcommand{\phiLBDP}{\phi_{\ell, \mathrm{BDP}}}
\newcommand{\thetaLBDP}{\theta_{\ell, \mathrm{BDP}}}
\newcommand{\OmegaLBDP}{\Omega_{\ell, \mathrm{BDP}}}

\usepackage{vmargin}
\setpapersize{A4}
\setmargins{2.5cm}       % left margin
{1.5cm}                        % superior margin
{16.5cm}                      % text width
{23.42cm}                    % text height
{10pt}                           % headlines' height
{1cm}                           % space between text and headlines
{0pt}                             % footnotes' height
{2cm}                           % space between text and footnote

\title{Can we measure Double DVCS at JLab and the EIC?}
\author[a]{K.~Deja\,\orcidlink{0000-0002-9083-2382}}
\author*[a]{V.~Mart\'inez-Fern\'andez\,\orcidlink{0000-0002-0581-7154}}
\author[b]{B.~Pire\,\orcidlink{0000-0003-4882-7800}}
\author[a]{P.~Sznajder\,\orcidlink{0000-0002-2684-803X}}
\author[a]{J.~Wagner\,\orcidlink{0000-0001-8335-7096}}

\affiliation[a]{National Centre for Nuclear Research (NCBJ),\\
  02-093 Warsaw, Poland}

\affiliation[b]{Centre de Physique Th\'eorique, CNRS, \'Ecole Polytechnique,\\
I.P. Paris, 91128 Palaiseau, France}

\emailAdd{katarzyna.deja@ncbj.gov.pl}
\emailAdd{victor.martinez-fernandez@ncbj.gov.pl}
\emailAdd{bernard.pire@polytechnique.edu}
\emailAdd{pawel.sznajder@ncbj.gov.pl}
\emailAdd{jakub.wagner@ncbj.gov.pl}

\abstract{Double deeply virtual Compton scattering (DDVCS) is a very precise tool for the nucleon tomography. Its measurement requires high luminosity electron beams and precise dedicated detectors, since its amplitude is quite small in the interesting kinematical domain where collinear QCD factorization allows the extraction of quark and gluon generalized parton distributions (GPDs). We analyze the prospects for its study in the JLab energy domain as well as in higher energy electron-ion colliders. Our results are very encouraging for various observables both with an unpolarized and polarized lepton beam. Using various realistic models for GPDs, we demonstrate that DDVCS measurements are indeed very sensitive to their behaviour. Implementing our lowest order cross-section formulae in the EpIC Monte Carlo generator, we estimate the expected number of interesting events. }

\FullConference{25th International Spin Physics Symposium (SPIN 2023)\\
 24-29 September 2023\\
 Durham, NC, USA\\}

\begin{document}
\maketitle

\section{Introduction}

\begin{figure}
    \centering
    \includegraphics[scale=0.8]{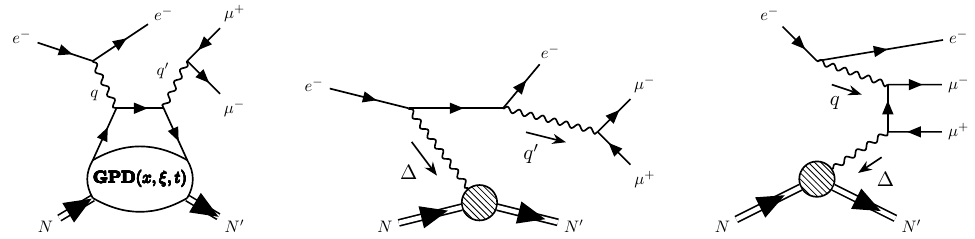}
    \caption{\scriptsize DDVCS (left) and Bethe-Heitler (BH, middle and right) lowest order Feynman diagrams for the electroproduction of a muon pair. Crossed diagrams are not shown.}
    \label{fig:ddvcs_and_bh}
\end{figure}

Double deeply virtual Compton scattering (DDVCS):
\begin{equation}
    \gamma^*(q) + N(p)  \to \gamma^*(q') + N'(p') \,,
    \label{reaction2}
\end{equation}
contributes to the exclusive electroproduction of a lepton pair,
\begin{equation}
    e(k) + N(p)  \to e'(k') + N'(p') + \mu^+(\ell_+) + \mu^-(\ell_-) \,,
    \label{reaction}
\end{equation}
which also receives contributions from  purely QED Bethe-Heitler processes (BH), as depicted in Fig.~\ref{fig:ddvcs_and_bh}.
Its amplitude does factorize \cite{Mueller:1998fv}  into perturbatively calculable coefficient functions and generalized parton distributions (GPDs) that unravel the three-dimensional structure of the nucleon \cite{Diehl:2003ny,Belitsky:2005qn} in the kinematical region where either $Q^2 = -q^2 = -(k-k')^2$ or $Q'^2 = q'^2 = (\ell_+ +\ell_-)^2$ is large, while the squared four-momentum transfer to the nucleon, $-t = -(p'-p)^2$ remains small. The importance of DDVCS has  already been emphasized some twenty years ago \cite{Belitsky:2002tf,guidal2003,Belitsky:2003fj}, but this process has never been measured so far; we stress in our studies \cite{Deja:2023ahc, Deja:2023tuc} that it should be studied  in the  future at both fixed-target facilities \cite{Chen:2014psa,Camsonne:2017yux,Zhao:2021zsm} and electron-ion colliders \cite{AbdulKhalek:2021gbh, Anderle:2021wcy}. 

\section{Kleiss-Stirling (KS) techniques}
In 1980s, Kleiss and Stirling \cite{Kleiss:1984dp,Kleiss:1985yh} envisaged an alternative way to use Feynman diagrams to compute cross-sections in quantum field theory. The main idea is to reduce the helicity amplitudes to complex numbers where spinors and Dirac-gamma matrices have been taken care of. For that purpose, two scalars are constructed as the building blocks:
\begin{align}
    s(a, b) & = \bar{u}(a,+)u(b,-) = (a^2 + ia^3)\sqrt{\frac{b^0 - b^1}{a^0 - a^1}} - (a\leftrightarrow b) = -s(b, a)\,, \label{sKS_def}\\
    t(a, b) & = \bar{u}(a,-)u(b,+) = [s(b, a)]^*\,, \label{tKS_def}
\end{align}
where $\pm$ stand for helicities and $a,b$ are light-like vectors. As mentioned before, these scalars are the building blocks of amplitudes as one can employ them to compute, for instance, the contraction of two currents
\begin{align}\label{function_f}
    f(\lambda, k_0, k_1; \lambda', k_2, k_3) = & \bar{u}(k_0,\lambda)\gamma^\mu u(k_1, \lambda)\bar{u}(k_2,\lambda')\gamma_\mu u(k_3,\lambda') \nonumber\\
    = & 2 [ s(k_2,k_1)t(k_0,k_3)\delta_{\lambda-}\delta_{\lambda'+} + t(k_2,k_1)s(k_0,k_3)\delta_{\lambda+}\delta_{\lambda'-} \nonumber\\
    & + s(k_2,k_0)t(k_1,k_3)\delta_{\lambda+}\delta_{\lambda'+} + t(k_2,k_0)s(k_1,k_3)\delta_{\lambda-}\delta_{\lambda'-} ]\,,
\end{align}
or the contraction of a current with a light-like vector $a$:
\begin{equation}\label{function_g}
    g(s, \ell, a, k) = \bar{u}(\ell, s)\slashed{a}u(k, s) 
    = \delta_{s+}s(\ell,a)t(a,k) + \delta_{s-}t(\ell,a)s(a,k)\,.
\end{equation}

\subsection{DDVCS subprocess {\`a} la KS}
For the case of DDVCS (left diagram in Fig.~\ref{fig:ddvcs_and_bh}), the Feynman amplitude at lowest (zeroth) order in the strong coupling constant and at leading twist can be written as:
\begin{equation}\label{iMddvcs}
    i\M_{\rm DDVCS} = \frac{-ie^4}{(Q^2-i0)(Q'^2+i0)}\left( i\M^{(V)}_{\rm DDVCS} + i\M^{(A)}_{\rm DDVCS} \right)\,,
\end{equation}
where $i\M^{(V)}_{\rm DDVCS}$ and $i\M^{(A)}_{\rm DDVCS}$ corresponds to the vector and the axial contributions to the amplitude, respectively, coming from the Compton tensor decomposition (to leading-twist accuracy)
\begin{equation}\label{comptonTensor}
    T^{\mu\nu}_{s_2s_1} =  -\frac{1}{2}g_\perp^{\mu\nu}\bar{u}(p',s_2)\left[ (\cffh + \cffe)\slashed{n} - \frac{\cffe}{M}\bp^+ \right] u(p, s_1) -\frac{i}{2}\epsilon^{\mu\nu}_\perp\bar{u}(p',s_2)\left[ \cffht\slashed{n} + \frac{\cffet}{2M}\Delta^+ \right]\gamma^5u(p,s_1)\,.
\end{equation}
By means of functions $f$ (\ref{function_f}) and $g$ (\ref{function_g}), they read:
\begin{align}\label{iMVddvcs}
    i\M^{(V)}_{\rm DDVCS} = & -\Bigg[ f(s_\ell, \ell_-, \ell_+; s, k', k) -  g(s_\ell,\ell_-,n^\star,\ell_+)g(s, k',n,k) - g(s_\ell,\ell_-,n,\ell_+)g(s, k',n^\star,k) \Bigg] \nonumber\\
    & \times \frac{1}{2} \Bigg[ (\cffh + \cffe) [ Y_{s_2s_1}g(+,r'_{s_2},n,r_{s_1}) + Z_{s_2s_1}g(-,r'_{-s_2},n,r_{-s_1}) ] - \frac{\cffe}{M} \Jt_{s_2s_1} \Bigg]\,,
\end{align}
and
\begin{equation}\label{iMAddvcs}
    i\M^{(A)}_{\rm DDVCS} = \frac{-i}{2} \epsilon^{\mu\nu}_\perp j_\mu(s_\ell,\ell_-,\ell_+)j_\nu(s, k', k)\left[ \cffht\Jofplus_{s_2s_1} + \cffet\frac{\Delta^+}{2M}\Jtfplus_{s_2s_1} \right]\,.
\end{equation}

In these sub-amplitudes, $M$ is the target mass, $f$ and $g$ are given in Eqs.~(\ref{function_f}) and (\ref{function_g}), and $\Jt$, $\Jofplus$, $\Jtfplus$, $Y$, $Z$ are combinations of scalars in Eqs.~(\ref{sKS_def}) and (\ref{tKS_def}) dependent on the spin and momentum of the target in its final ($s_2, p'$) and initial ($s_1, p$) states.  $j_\mu$ stands for the lepton current. Finally, $n$ and $n^\star$ are vectors defining the ``plus'' and ``minus'' light-cone directions as considered in \cite{Belitsky_2000, Belitsky:2003fj, Deja:2023ahc}. This formulation, including BH contributions, was numerically validated against the DVCS and TCS limits \cite{Deja:2023ahc, Deja:2023tuc}.

The results for longitudinally and transversely polarized targets can also be accessed within the Kleiss-Stirling approach. Thus far, hadron polarization denoted with index $s_1$ corresponds to the values $\pm$ for helicity with respect to the three-vector component of $s^\mu = (r_1^\mu - r_2^\mu)/M$\,, where $r_1,r_2$ are two light-like vectors such that $p = r_1+r_2$\,. The relation between the quantization of helicity-spinors in direction $\vec{s}$ (denoted by $s_1=\pm$) and in another direction defined by three-vector\begin{NoHyper}\footnote{Angles $\phiS$ and $\thetaS$ are the azimuthal and polar orientations of $\vec{S}$ with respect to the target rest frame TRF-II, vid.~\cite{Deja:2023ahc}.}\end{NoHyper} $\vec{S} = (\sin\thetaS\cos\phiS,\sin\thetaS\sin\phiS,\cos\thetaS)$ is given by the relation $u'(p, h_1) = F_{h_1 +} u(p,+) + F_{h_1 -} u(p,-)$\,, where the matrix $F$ has been defined in~\cite{Deja:2023ahc}. Therefore, the correspondence between the amplitudes above, where the target is polarized in the direction $\vec{s}$ (index $s_1$), and the ones with a target polarized with respect to $\vec{S}$ (index $h_1$) is nothing but $i\M(s_2,h_1) = F_{h_1+}i\M(s_2,s_1=+) + F_{h_1-}i\M(s_2,s_1=-)$\,, where $s_2$ is the helicity of the final-state proton. For further details, cf.~Ref.~\cite{Deja:2023ahc}.

\section{DDVCS observables}
Let us now present selected DDVCS observables in the kinematics of current and future experiments, showing that the GPD model dependence can also be addressed. For this purpose we use the GK \cite{Goloskokov_2007, Goloskokov_2007_2}, VGG \cite{guichon1998, guichon1999, Goeke_2001, guidal2005} and MMS \cite{Mezrag:2013mya} GPD models implemented in the PARTONS software framework~\cite{Berthou:2015oaw}. For angles referring to the produced lepton pair, denoted with subscript $\ell$, we make use of the BDP frame \cite{Berger:2001xd} which is common in TCS studies. 

The selected observables are unpolarized differential cross-sections (right and left arrows stand for positive and negative helicity of the incoming electron beam, respectively):
\begin{equation}
    \sigma_{UU}(\phiLBDP) = \int_0^{2\pi} d\phi\int_{\pi/4}^{3\pi/4}d\thetaLBDP\ \sin\thetaLBDP \scriptstyle{\left( \frac{d^7\sigma^{\rightarrow}}{dx_B dQ^2 dQ'^2 d|t| d\phi d\OmegaLBDP} + \frac{d^7\sigma^{\leftarrow}}{dx_B dQ^2 dQ'^2 d|t| d\phi d\OmegaLBDP} \right)}\,,
    \label{eq:ass1}
\end{equation}
and their cosine components: $\sigma_{UU}^{\cos(n \phiLBDP)}(\phiLBDP)$\,. We also consider asymmetries for a longitudinally polarized electron beam $A_{LU}(\phiLBDP) = \Delta\sigma_{LU}(\phiLBDP) / \sigma_{UU}(\phiLBDP)$\,, where $\Delta\sigma_{LU}$ corresponds to the same integral as (\ref{eq:ass1}) up to the change $(d^7\sigma^{\rightarrow} + d^7\sigma^{\leftarrow}) \rightarrow (d^7\sigma^{\rightarrow} - d^7\sigma^{\leftarrow})$.

\begin{figure}[htb]
    \centering
    \includegraphics[width=0.24\textwidth]{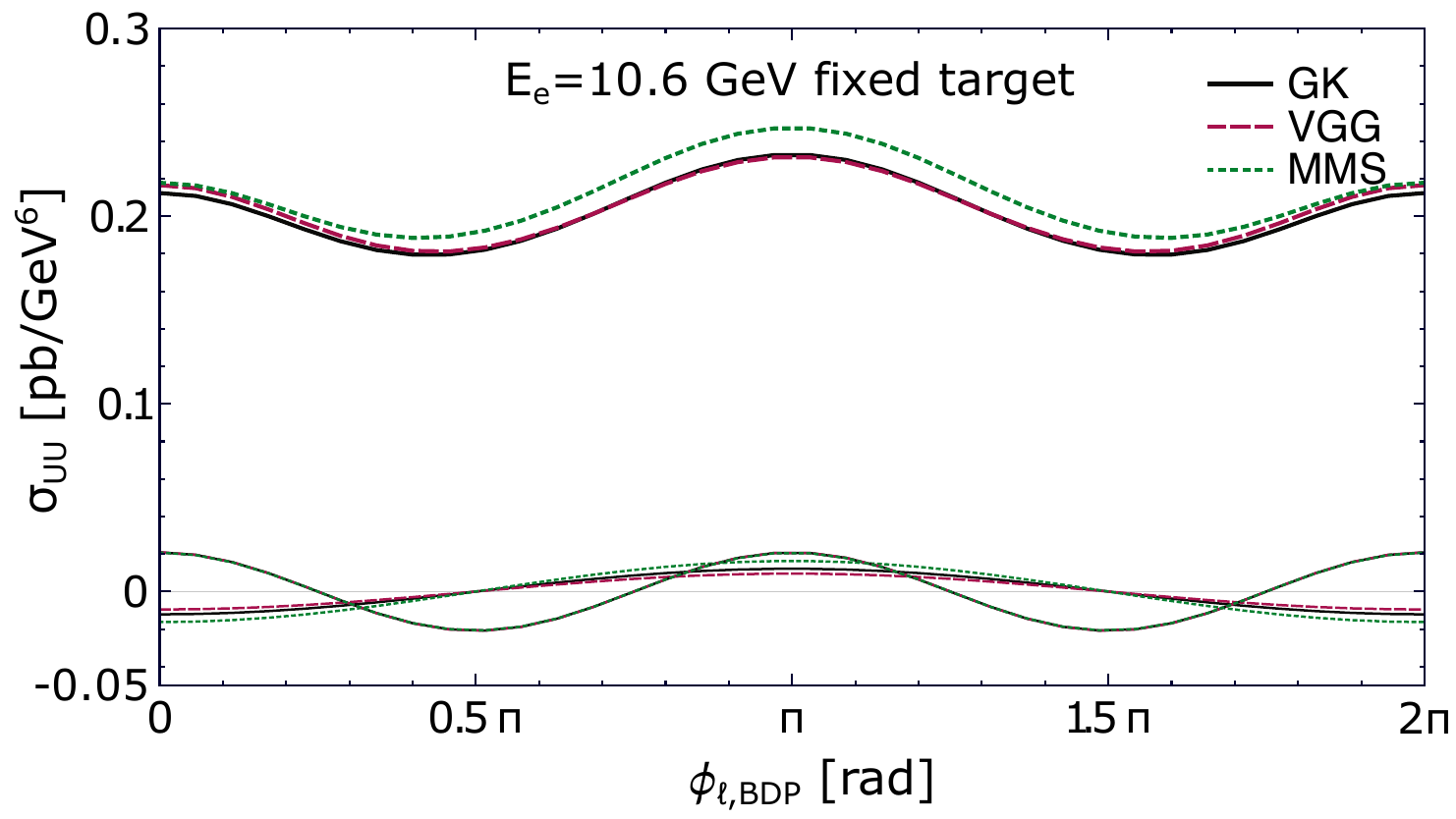}
    \includegraphics[width=0.24\textwidth]{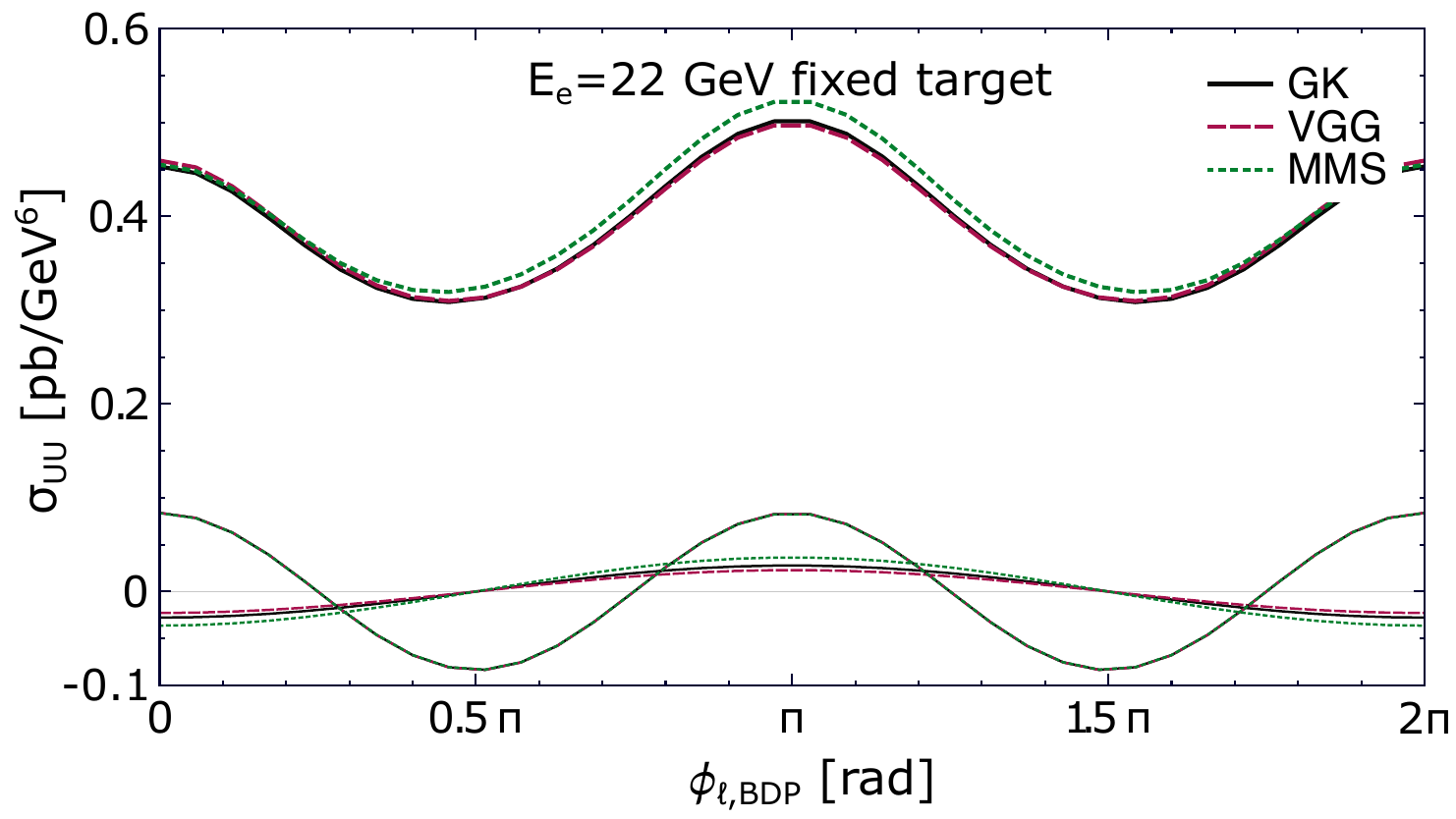}
    \includegraphics[width=0.24\textwidth]{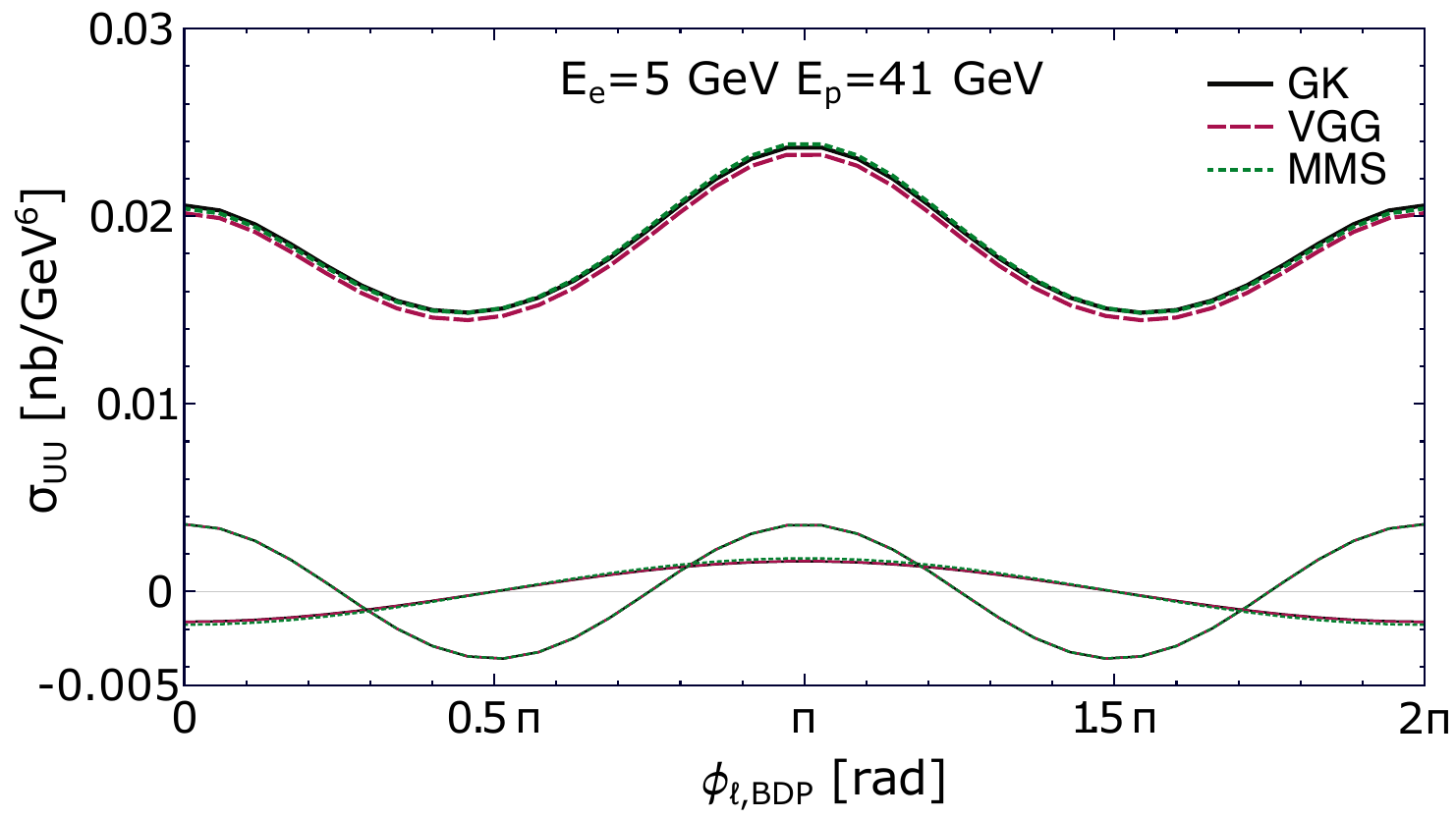}
    \includegraphics[width=0.24\textwidth]{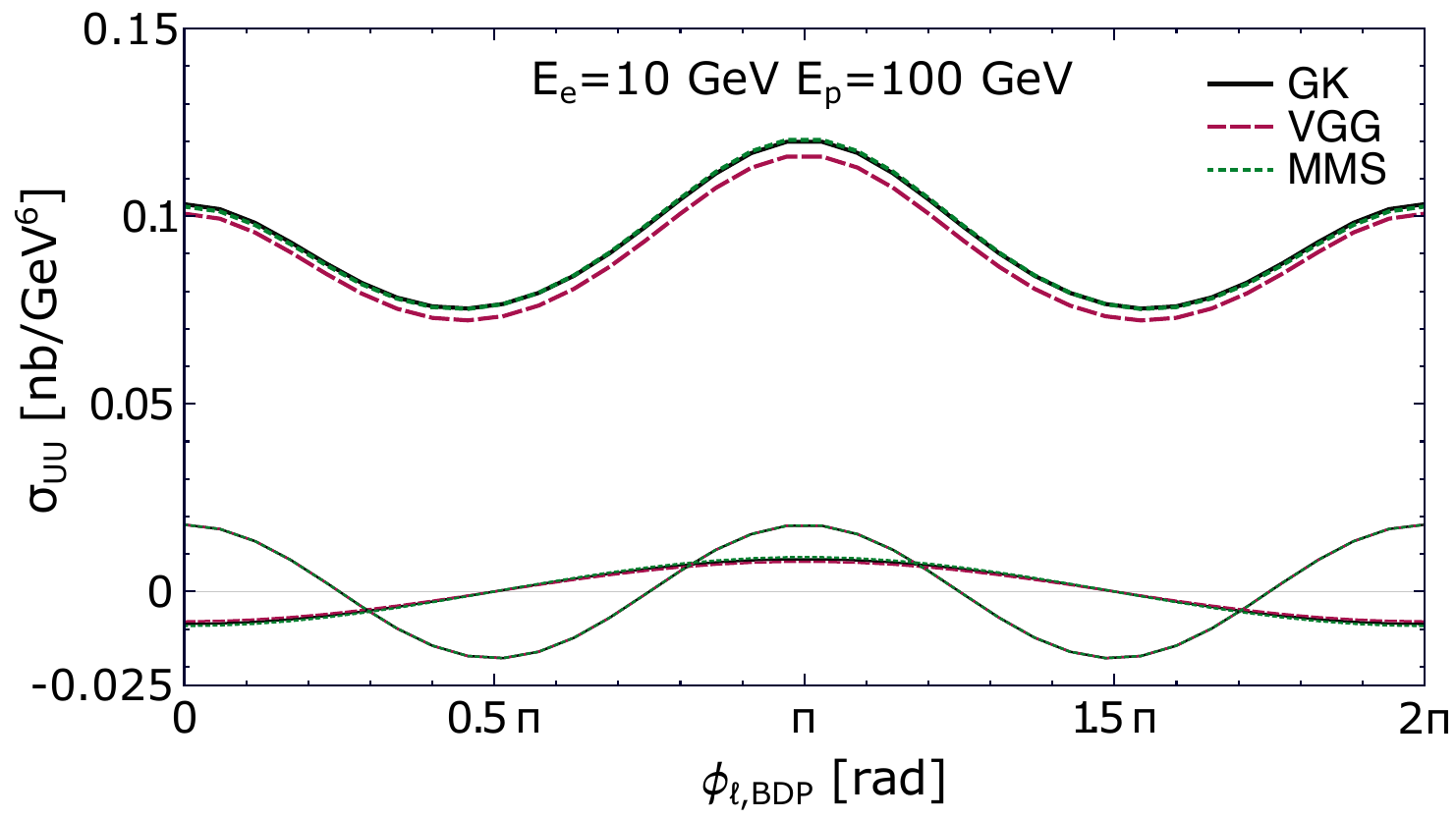}
    \caption{\scriptsize Unpolarized cross-section, $\sigma_{UU}(\phiLBDP)$, and its  $\sigma_{UU}^{\cos\phiLBDP}(\phiLBDP)$ and $\sigma_{UU}^{\cos2\phiLBDP}(\phiLBDP)$ components for beam energies specified in the plots and extra kinematic conditions given in text. From left to right: JLab12, JLab20+, EIC 5$\times$41 and EIC 10$\times$100.}
    \label{fig:denominators}
\end{figure}

\begin{figure}[htb]
    \centering
    \includegraphics[width=0.24\textwidth]{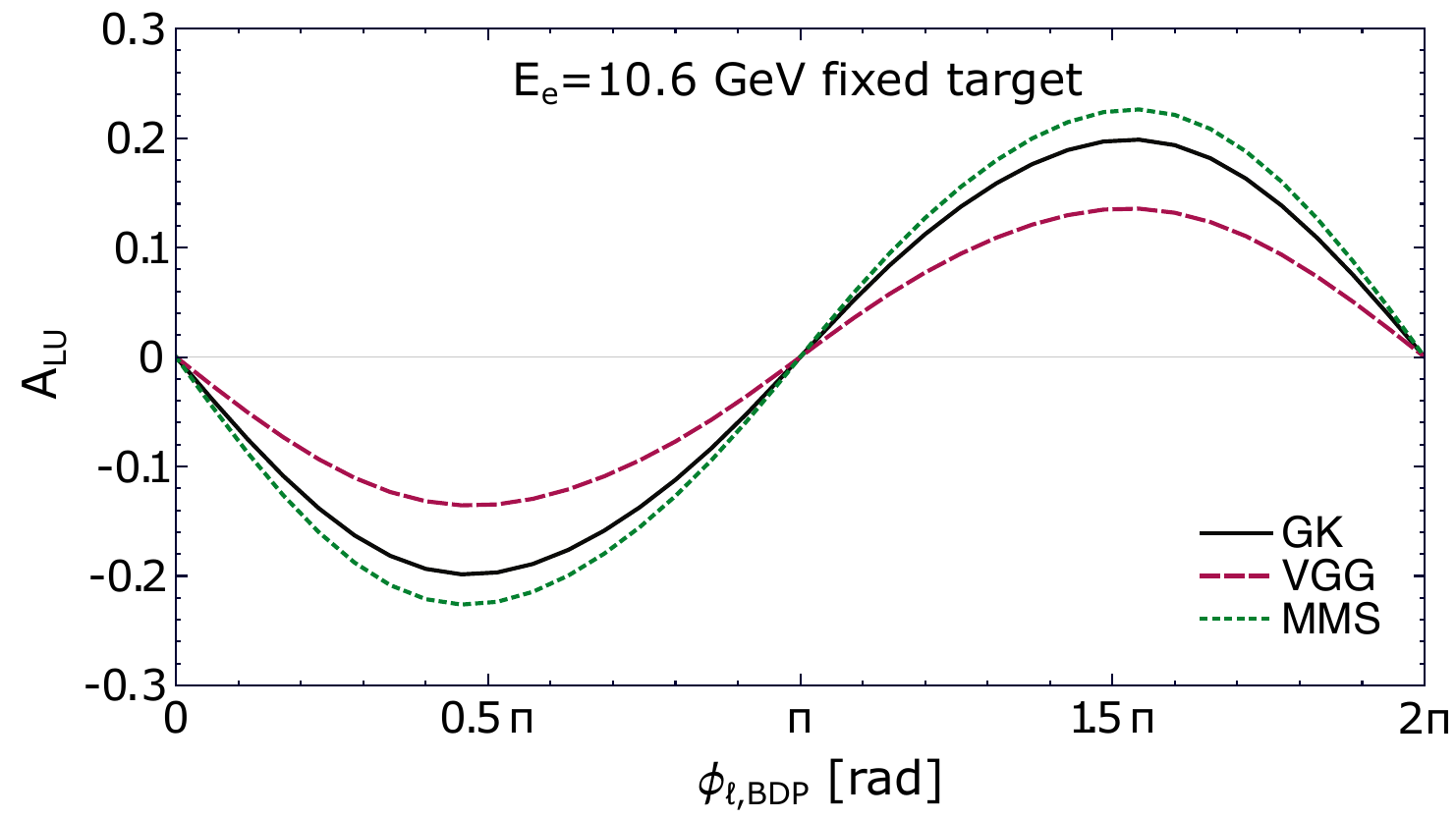}
    \includegraphics[width=0.24\textwidth]{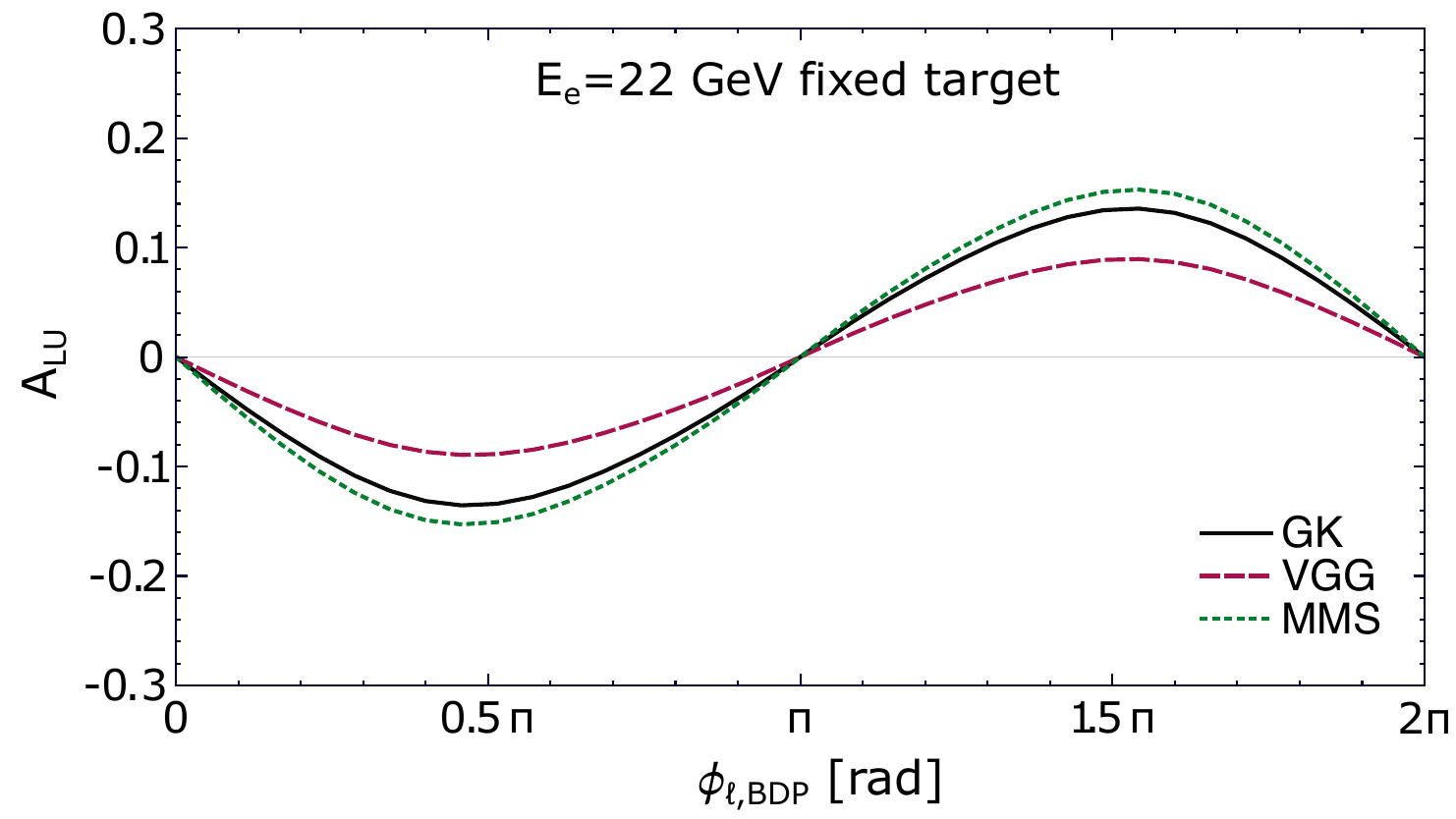}
    \includegraphics[width=0.24\textwidth]{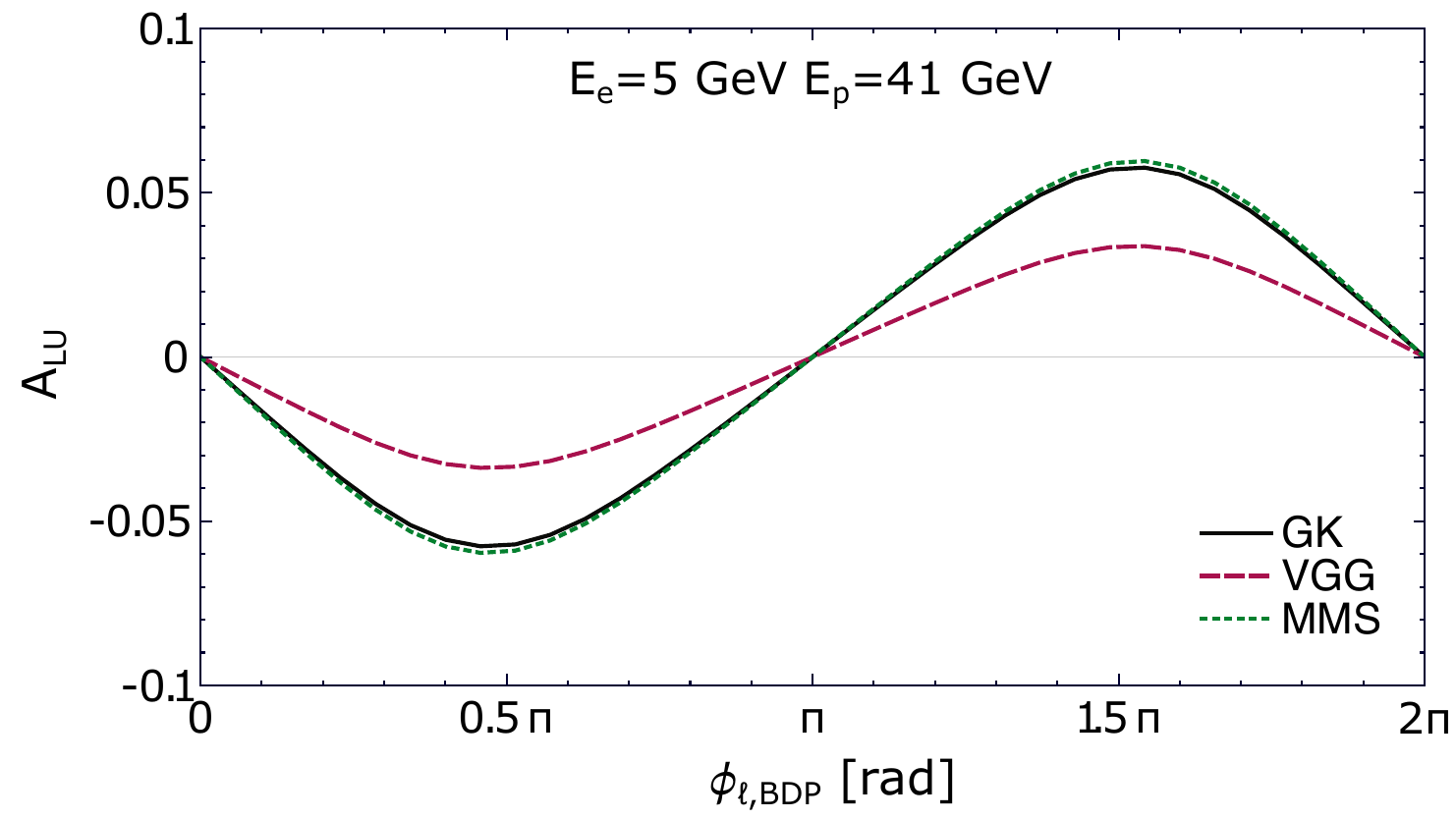}
    \includegraphics[width=0.24\textwidth]{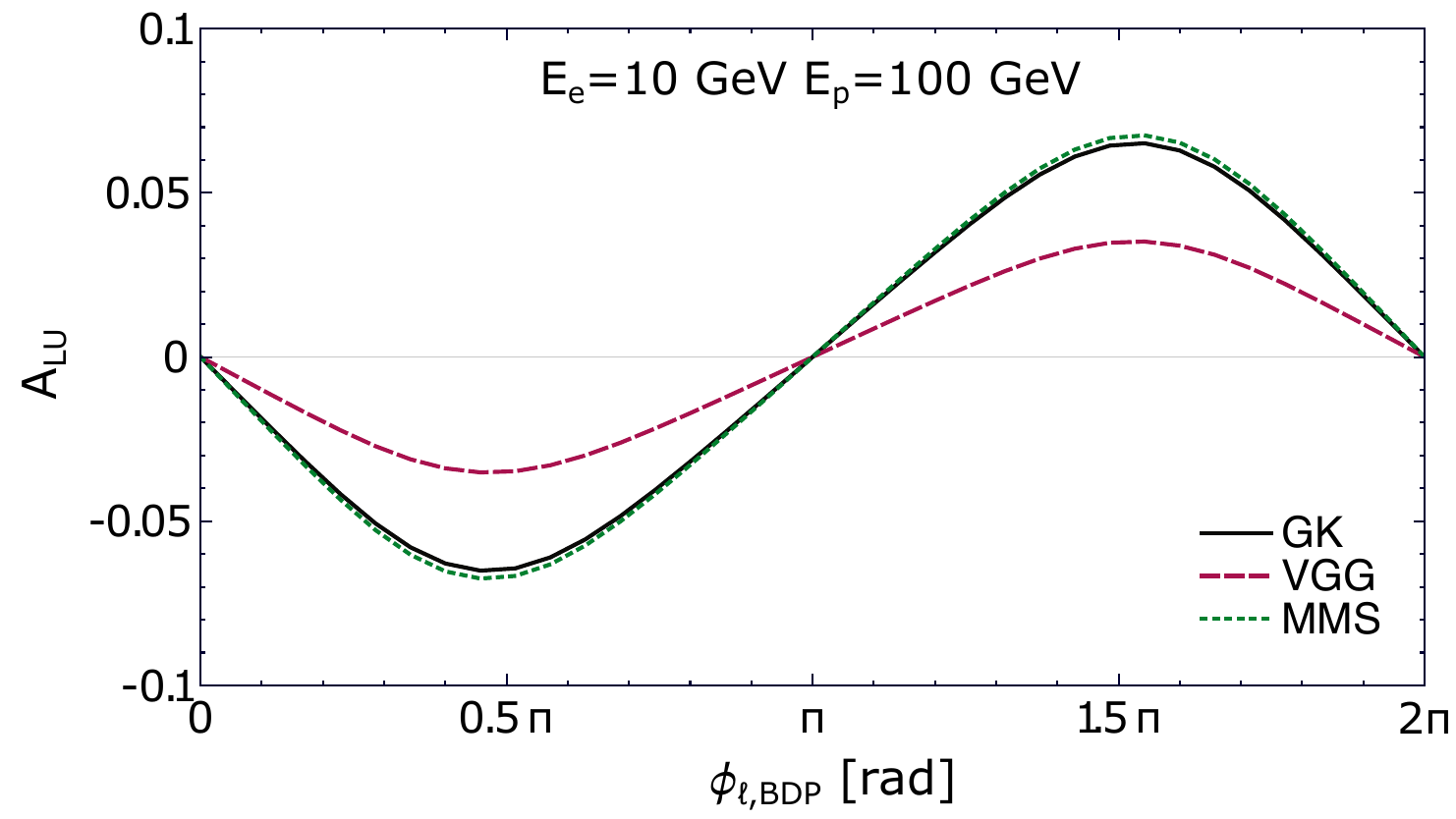}
    \caption{\scriptsize Asymmetry $A_{LU}(\phiLBDP)$ for beam energies specified in the plots and extra kinematic conditions given in text. From left to right: JLab12, JLab20+, EIC 5$\times$41 and EIC 10$\times$100.}
    \label{fig:asymetries}
\end{figure}

The predictions for JLab12, JLab20+ and EIC experiments (for two configurations of beam energies) are shown in Fig.~\ref{fig:denominators} for unpolarized cross-sections and their cosine components, and in Fig.~\ref{fig:asymetries} for the lepton beam helicity asymmetry $A_{LU}$. We choose the kinematical point, $t = -0.2$ GeV$^2, y = 0.5$  for JLab12, and $t = -0.2$ GeV$^2, y = 0.3$ for JLab20+. For both configurations of EIC, we choose $t = -0.1$ GeV$^2$, $y = 0.15$. In all experiments, $Q^2 = 0.6$ GeV$^2$ and $Q^{\prime 2} = 2.5$ GeV$^2$.


\section{Monte Carlo study}

\begin{figure}[htb]
    \centering
    \includegraphics[width=0.24\textwidth]{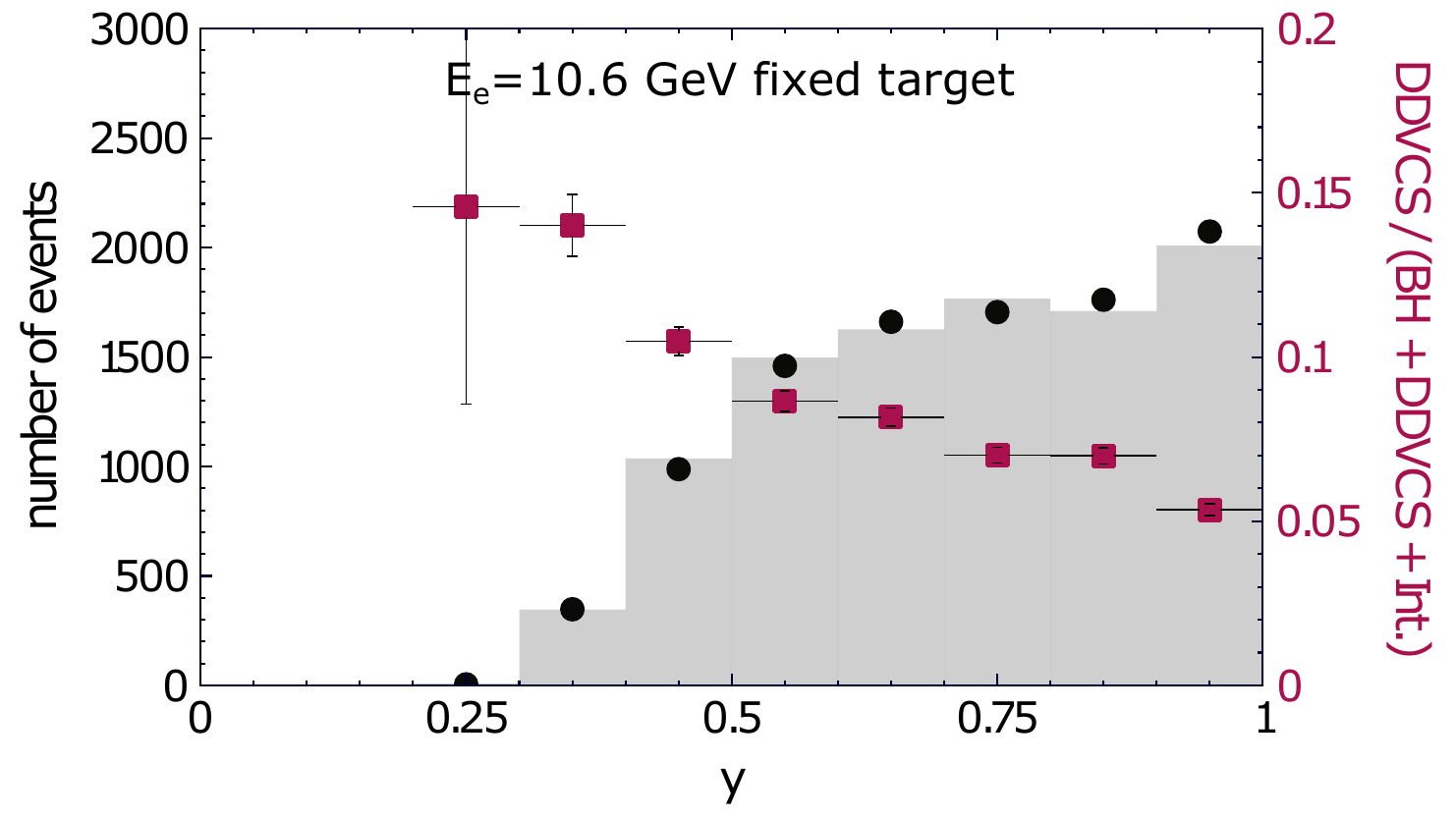}
    \includegraphics[width=0.24\textwidth]{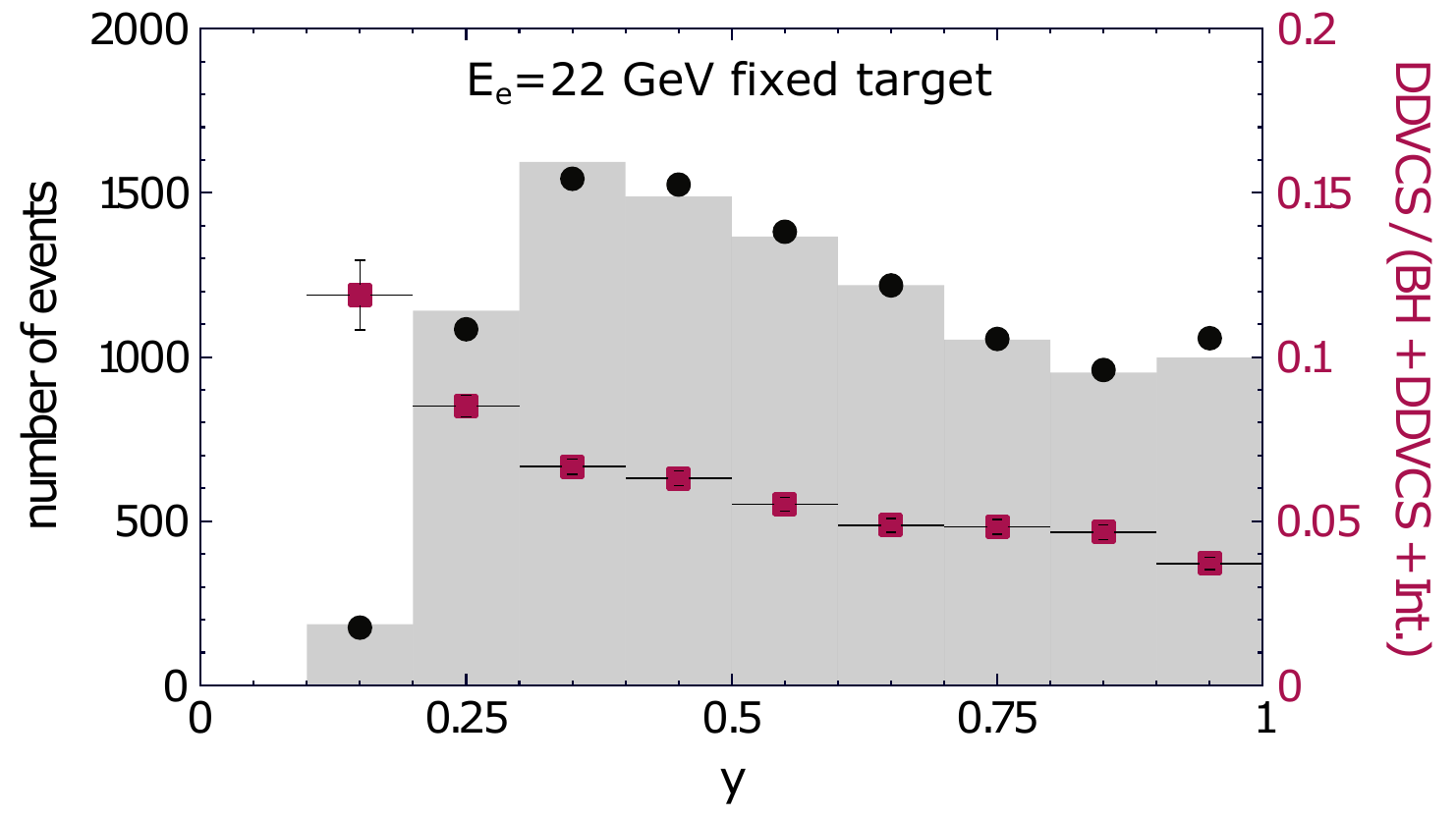}
    \includegraphics[width=0.24\textwidth]{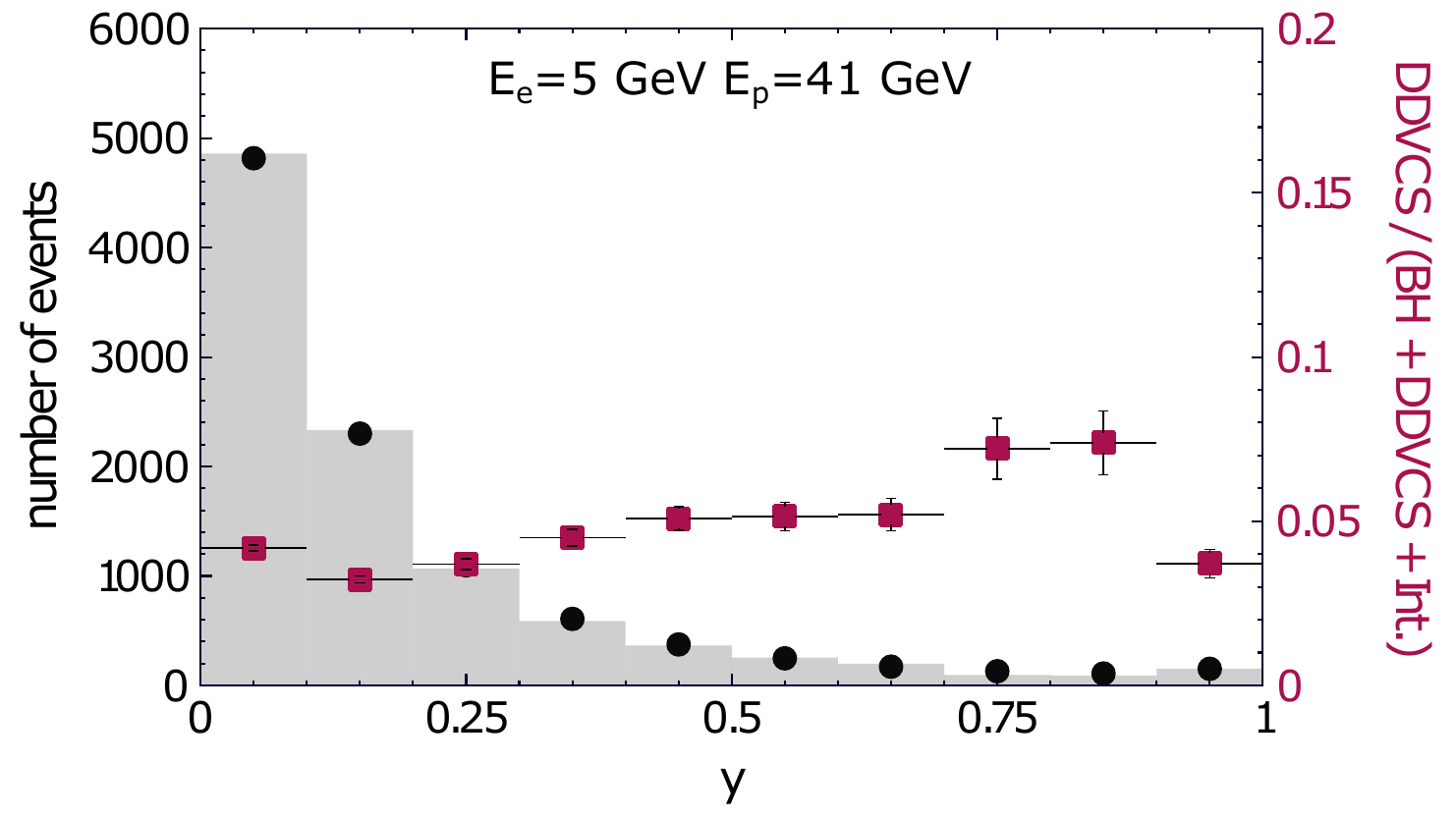}
    \includegraphics[width=0.24\textwidth]{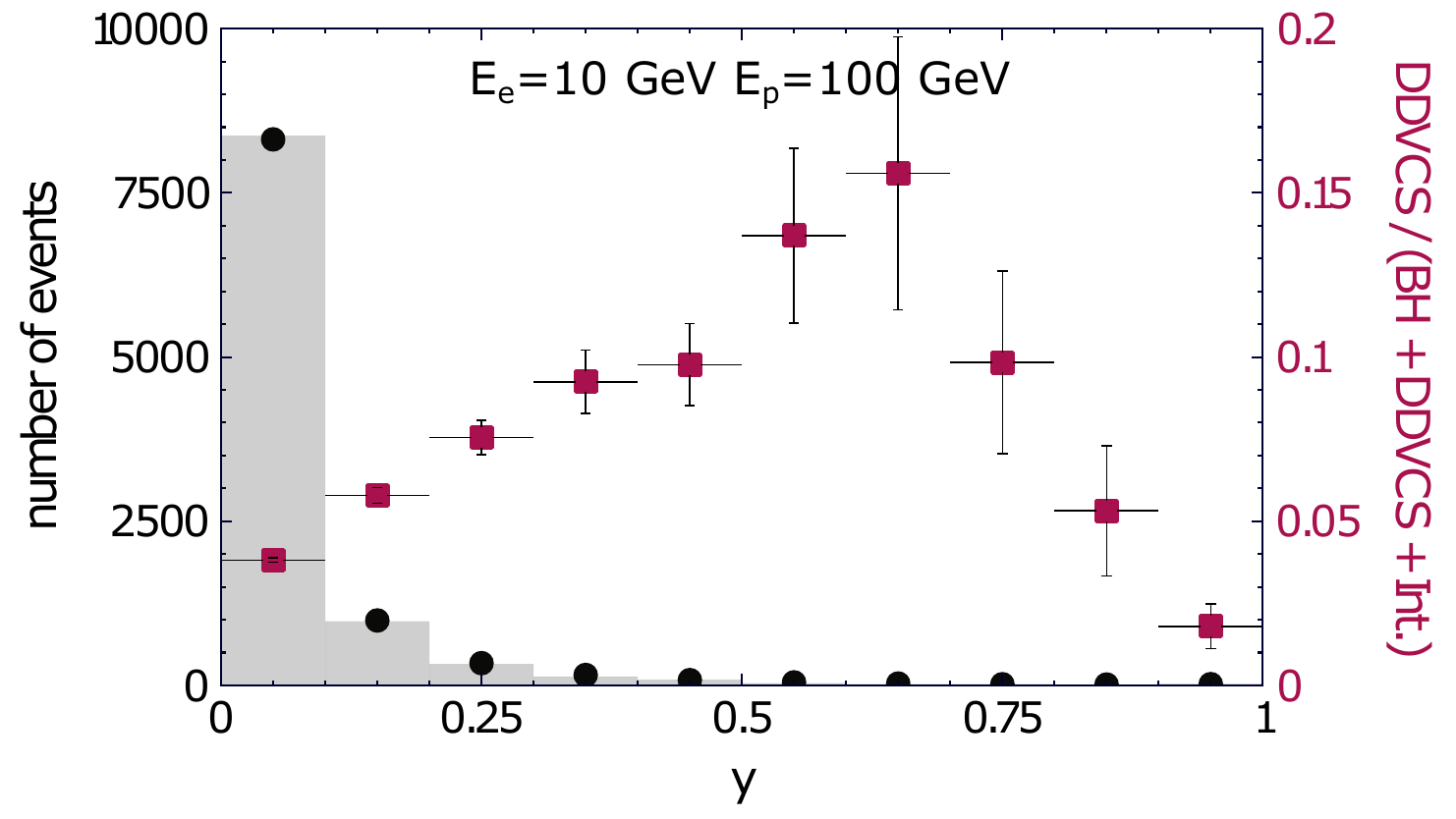}
    \caption{\scriptsize Distributions of Monte Carlo events as a function of the inelasticity variable $y$. Each distribution is populated by 10000 events generated for the beam energies specified in the plots. Extra kinematics indicated in the text. From left to right: JLab12, JLab20+, EIC 5$\times$41 and EIC 10$\times$100.}
    \label{fig:MCHist}
\end{figure}

\begin{figure}
    \centering
    \includegraphics[width=0.24\textwidth]{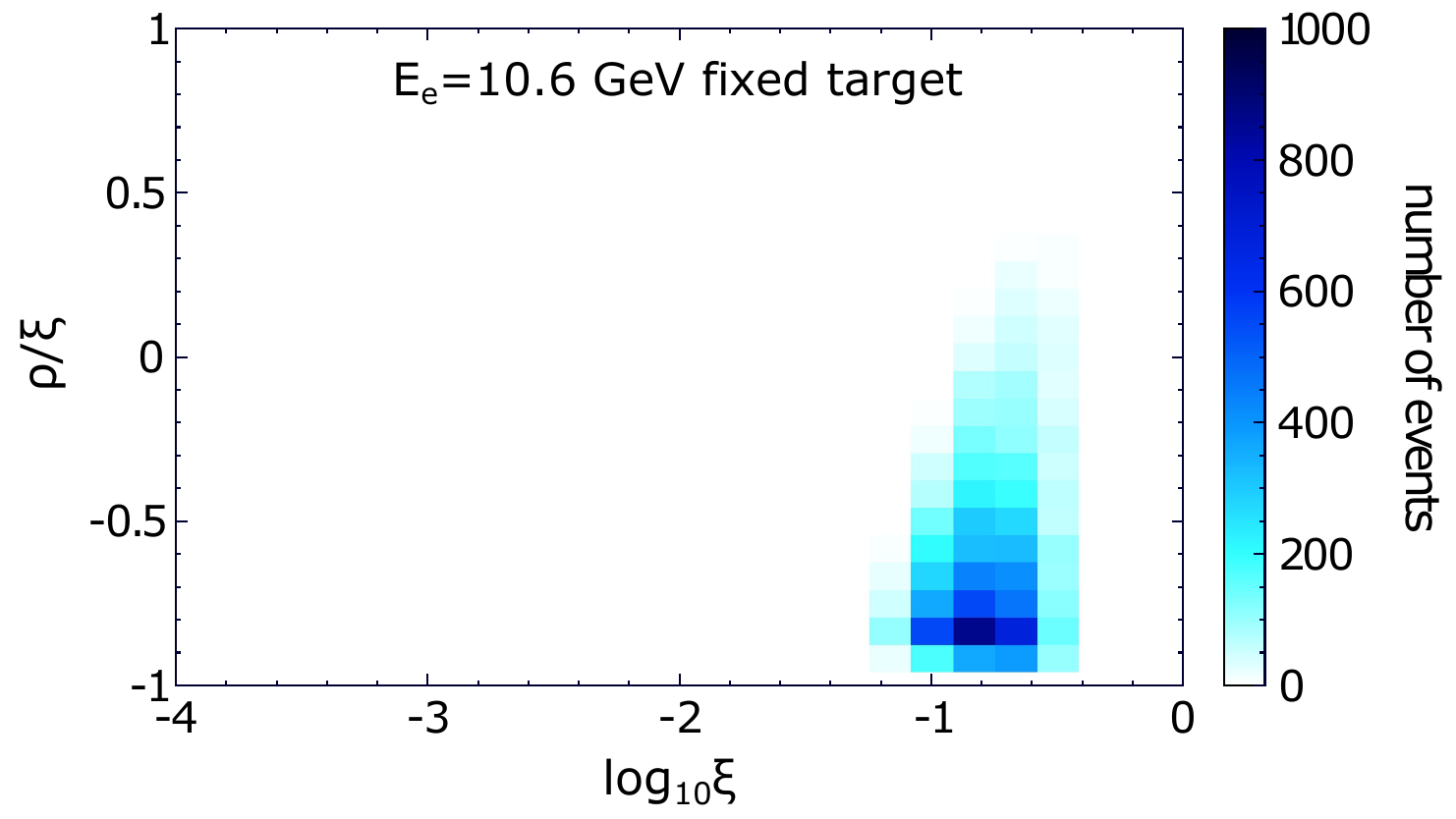}
    \includegraphics[width=0.24\textwidth]{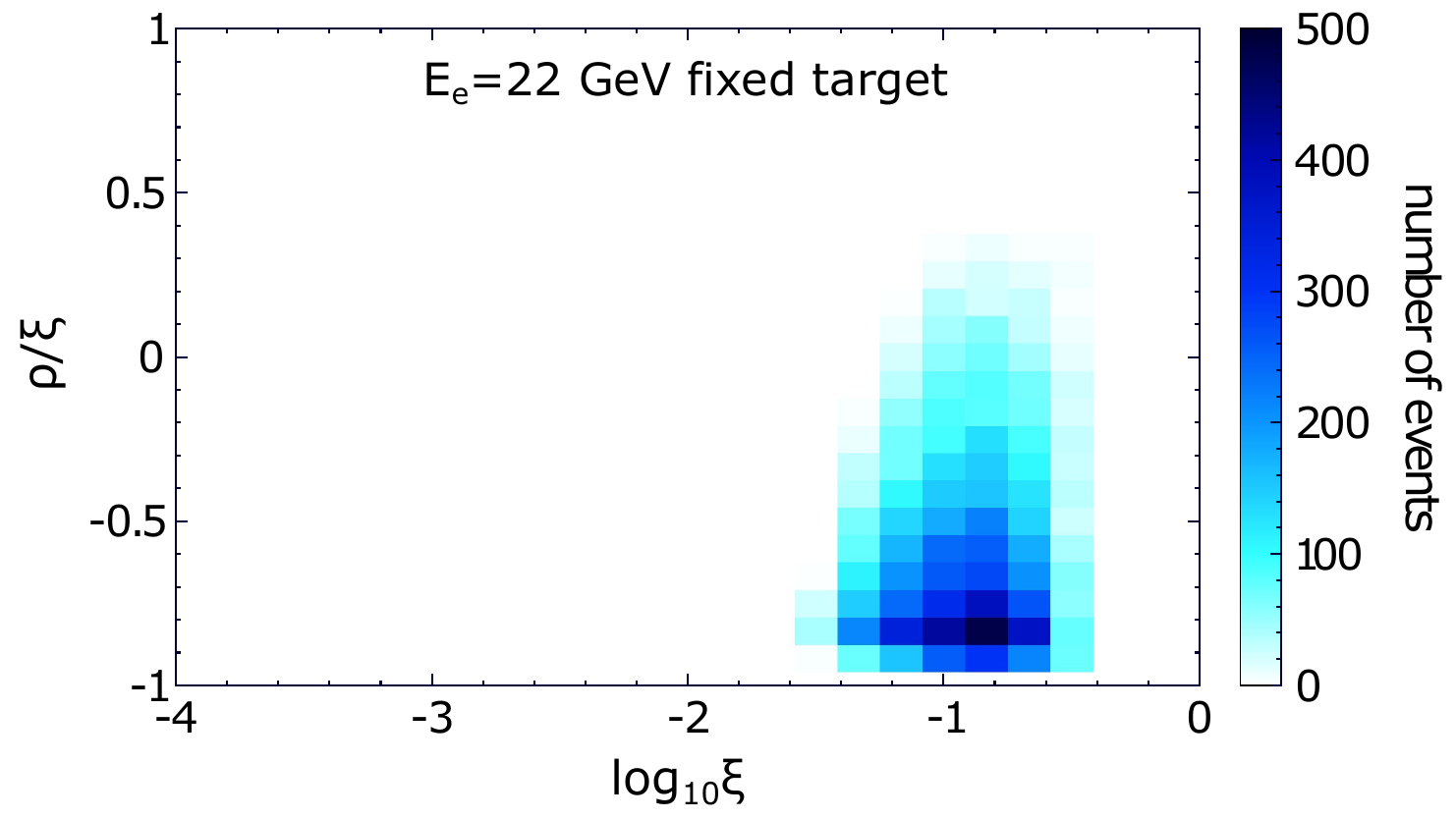}
    \includegraphics[width=0.24\textwidth]{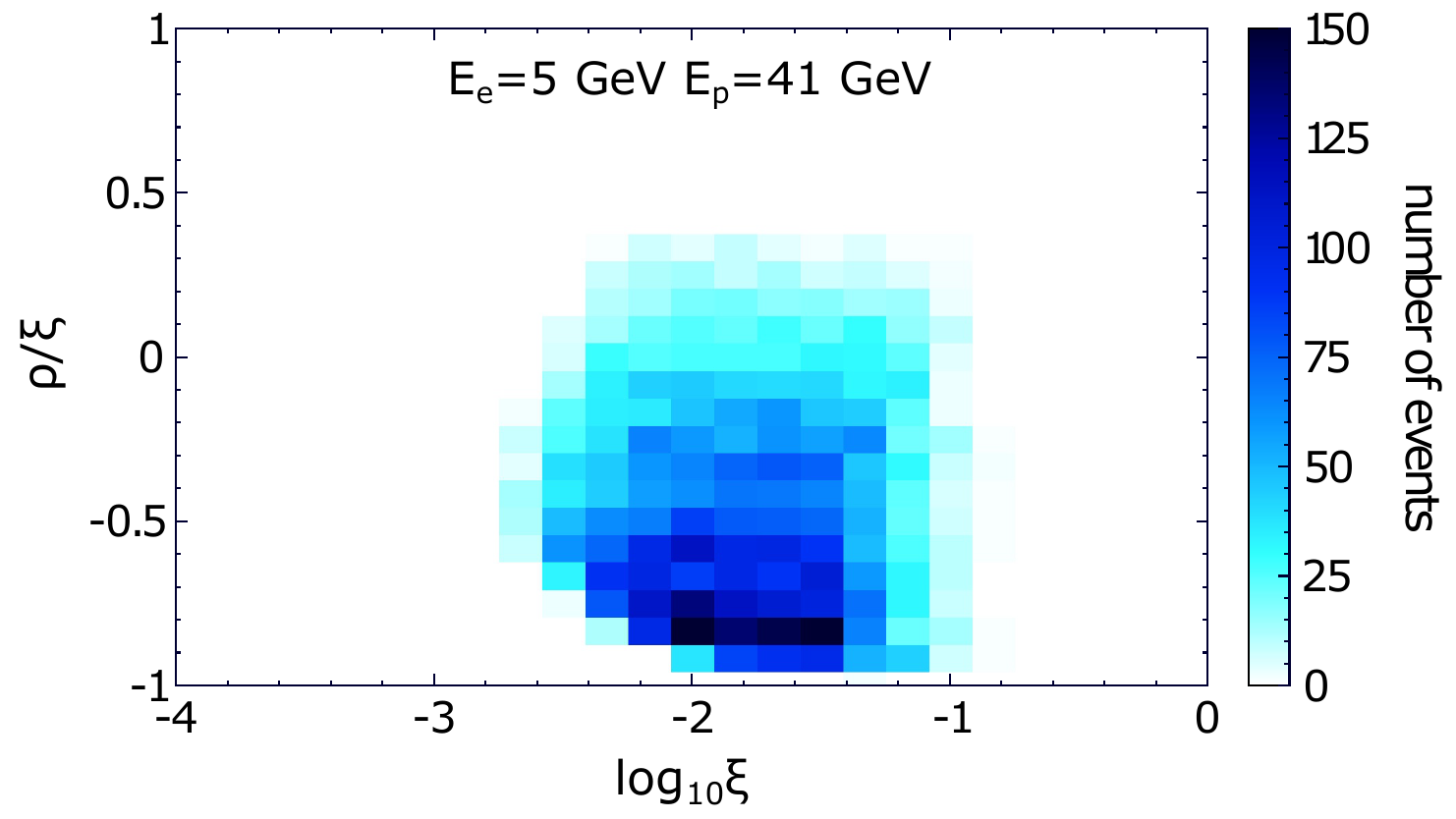}
    \includegraphics[width=0.24\textwidth]{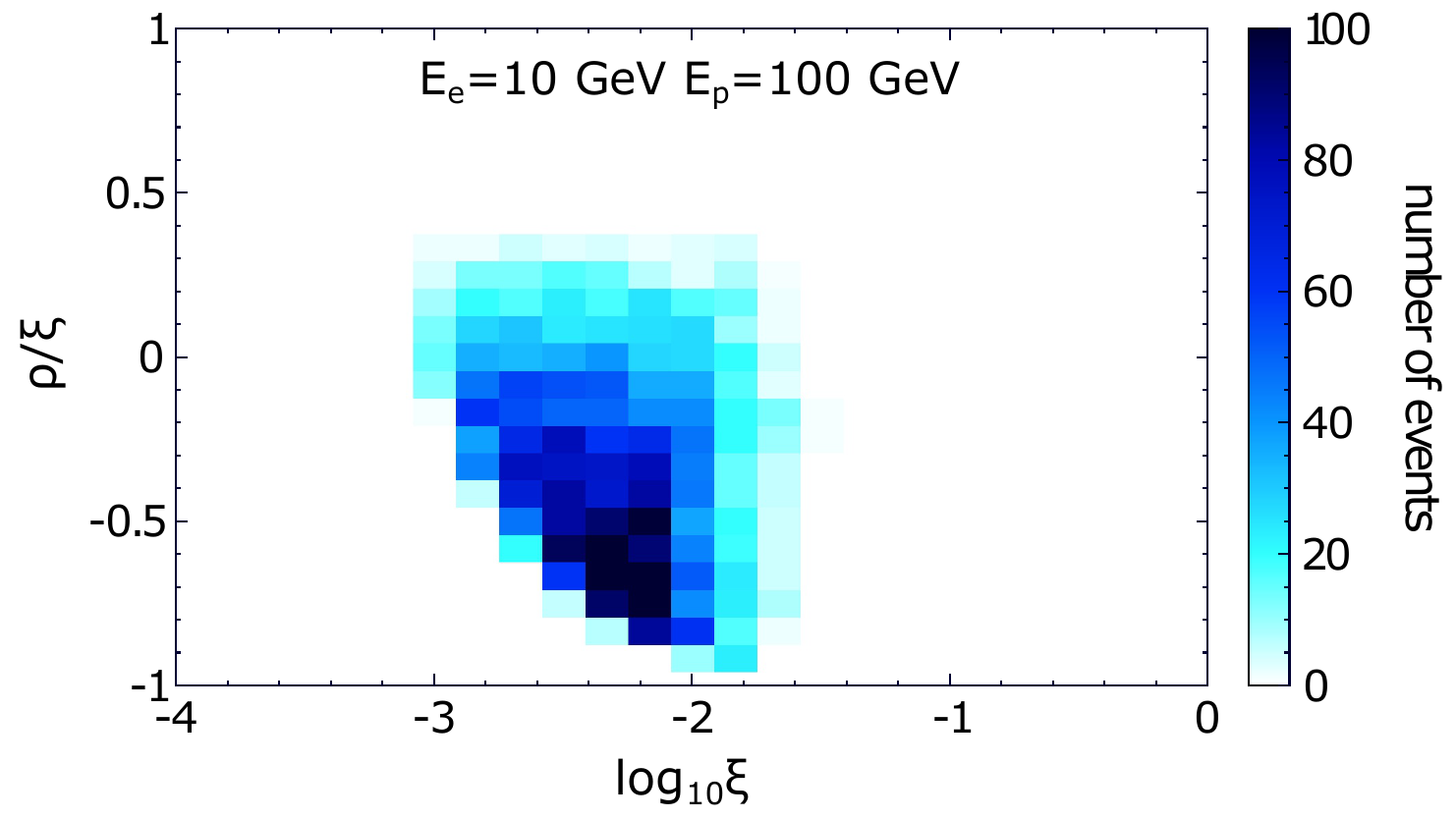}
    \caption{\scriptsize Distribution of Monte Carlo events as a function of the skewness variable $\xi$ and the relative value of generalized Bj\"orken variable $\rho$. Each distribution is populated by 10000 events generated for the DDVCS sub-process at beam energies specified in the plot. Extra kinematical conditions, including cuts on the $y$ variable, are specified in the text.}
    \label{fig:MCxiVsRho}
\end{figure}

The cross-section formulae we obtained have been implemented in the open-source PARTONS framework~\cite{Berthou:2015oaw}. The implementation of DDVCS in the EpIC Monte Carlo (MC) generator \cite{Aschenauer:2022aeb} has followed, making our work directly applicable for experimental analysis. Our results do not include any acceptances or detector response. 

The distribution of MC events as a function of the inelasticity variable $y = (E_e-E'_e)/E_e$ is shown in Fig.~\ref{fig:MCHist}. Here, $E_e$ and $E'_e$ represent the energies of the incoming and scattered electron beam, respectively. The total cross-section for the total scattering is given in Tab.~\ref{tab:MCCS}. Kinematical cuts for the integration are: $y \in (0, 1)$, $Q^2 \in (0.15, 5)\ \mathrm{GeV}^2$, $Q^{\prime 2} \in (2.25, 9)\ \mathrm{GeV}^2$, $\phi, \phiL \in(0.1, 2\pi - 0.1)$ rad, $\thetaL \in (\pi/4, 3\pi/4)$ rad and $|t|\in(0.1, 0.8)$ GeV$^2$ for JLab and $(0.05, 1)$ GeV$^2$ for EIC. In this table we also specify the integrated luminosity needed to record 10000 events presented in Fig.~\ref{fig:MCHist}, and the fraction of events recovered after accounting for the lower cut in $y$, namely $y_{\rm min}$ in the table, which is related to the experimental energy resolution. In Fig.~\ref{fig:MCHist} we also show the expected number of events, coming from a direct seven-fold integration of cross-section (black dots), compared to the MC samples (grey bands) generated by EpIC. The smallness of the fraction of pure DDVCS contribution (red boxes) suggests considering observables dependent on the interference between the DDVCS and BH subprocesses. 

Additionally, Fig.~\ref{fig:MCxiVsRho} depicts the distribution of DDVCS over the $\rho,\xi$ phase-space, showing clear access to the ERBL region, fundamental for the extraction of GPDs. In such plots, we make use of the same constraints detailed above.

\begin{table}[htb]
\centering
\scalebox{0.9}{\begin{tabular}{lcccccc}
\hline\hline
& & & & & \\[-12pt]
Experiment & Beam energies & Range of $|t|$ & $\sigma \rvert_{0<y<1}$ & $\mathcal{L}^{10\mathrm{k}}\rvert_{0<y<1}$ & $y_{\mathrm{min}}$ & $\sigma \rvert_{y_{\mathrm{min}} < y < 1} / \sigma \rvert_{0<y<1}$ \\ 
& $[\mathrm{GeV}]$ & $[\mathrm{GeV}^2]$ & $[\mathrm{pb}]$ & $[\mathrm{fb}^{-1}]$ & & \\[5pt] 
JLab12 & $E_{e} = 10.6$, $E_p = M$ & $(0.1, 0.8)$ & $0.14$ & $70$ & $0.1$ & $1$\\
JLab20+ & $E_{e} = 22$, $E_p = M$ & $(0.1, 0.8)$ & $0.46$ & $22$ & $0.1$ & $1$\\ 
EIC & $E_{e} = 5$, $E_p = 41$ & $(0.05, 1)$ & $3.9$ & $2.6$ & $0.05$ & $0.73$\\
EIC & $E_{e} = 10$, $E_p = 100$ & $(0.05, 1)$ & $4.7$ & $2.1$ & $0.05$ & $0.32$ \\
[3pt]
\hline\hline
\end{tabular}}
\caption{\scriptsize Total cross-section for electroproduction of a muon pair, $\sigma \rvert_{0<y<1}$, obtained for the beam energies indicated in each plot, $y\in(0, 1)$ and extra kinematics indicated in the text. Corresponding integrated luminosity required to obtain 10000 events is denoted by $\mathcal{L}^{10\mathrm{k}}\rvert_{0<y<1}$. Fraction of events left after restricting the range of $y$ to $(y_{\mathrm{min}}, 1)$ is given in the last column.}
\label{tab:MCCS}
\end{table}

As a final remark, we conclude that asymmetries at LO are of order 15-20\% for JLab and 3-7\% for EIC, large enough for DDVCS to be considered as a relevant part of GPD programms in current and future experimental facilities. The inclusion of NLO corrections \cite{Pire:2011st} should not change this conclusion. Since the DDVCS cross section is mostly accessible at moderate values of $Q^2, Q'^2$ (compared to $t$ and nucleon mass), it is important to estimate the effects of kinematic higher-twist corrections in the line of recent works on the DVCS process \cite{Braun:2014sta,Braun:2022qly} based on the conformal operator-product expansion developed in \cite{Braun:2020zjm}. We shall soon explore these contributions with application to imaging through nucleon tomography. This, in itself, requires knowledge over a sizable range of $t$\,.\\\vspace{-2mm}

\scriptsize{{\bf Acknowledgements.} This work is partly supported by the COPIN-IN2P3 and by the European Union’s Horizon 2020 research and innovation programme under grant agreement No 824093. The works of V.M.F.~are supported by PRELUDIUM grant 2021/41/N/ST2/00310 of the Polish National Science Centre (NCN).}

\addcontentsline{toc}{section}{References}
\bibliographystyle{JHEP}
\bibliography{references}

\end{document}